\documentstyle[preprint,aps]{revtex}
\widetext
\draft
\tighten

\begin{document}
\newcommand{\bfbeta}{\mbox{\boldmath$\beta$}}
\newcommand{\bfeta}{\mbox{\boldmath$\eta$}}
\newcommand{\wpart}{\hat{\partial}}
\newcommand{\tpr}{t^{\prime}}


\title{BROWNSTEIN'S WHOLE-PARTIAL DERIVATIVES: THE CASE OF THE LORENTZ
GAUGE}

\bigskip

\author{{\bf A. Chubykalo, A. Espinoza and R. Flores-Alvarado}}
\address{{\rm Escuela de F\'{\i}sica, Universidad Aut\'onoma de
Zacatecas \\
Apartado Postal C-580,\\
Zacatecas 98068, ZAC., M\'exico\\
e-mail: andrew\_chubykalo@att.net.mx}}


\maketitle

\baselineskip 7mm

\bigskip

\begin{center}
Received $\;\;\;\;\;\;\;\;\;\;\;\;\;\;\;\;\;\;\;\;\;\;\;$ 2004
\end{center}

\begin{abstract}
In this brief note we show that the usual Lorentz gauge is not satisfied
by the Lienard-Wiechert potentials, then, using Brownstein's concept of
{\it whole-partial derivatives} we introduce the generalized expression
for the Lorentz gauge showing that it is satisfied by the LW-potentials.
\end{abstract}

$$$$$$$$


\clearpage

\section{Introduction}

In the paper of 1997 [1] a way was proposed  to generalize the Maxwell
equations using {\it total (whole)} derivatives instead of partial ones.
However, the concept of total derivative in some special cases was
criticized and refined by K. R. Brownstein [2] which introduced and
rationalized so-called ``{\it whole-partial}" derivative [3]. Some related
ideas were discussed by A. E. Chubykalo and R. A. Flores [4] trying to
give a theoretical rationale for the {\it correct} use of total derivative
concept in mathematical analysis.

In this paper we shall show that the Lorentz gauge as written in the form:

\begin{equation}
\sum_{i=1}^{3}\frac{\partial A_{i}}{\partial x_{i}}+\frac{1}{c}\frac{
\partial \varphi }{\partial t}=0\qquad{\rm or}\qquad
\sum_{i=1}^3\varphi\frac{\partial\beta_i}{\partial x_i}+
\sum_{i=1}^3\beta_i\frac{\partial\varphi}{\partial x_i}+
\frac{1}{c}\frac{\partial\varphi}{\partial t}=0
\end{equation}
is not satisfied when $A_{i}$ and $\varphi$ are given by the
Li\'enard-Wiechert (LW) potentials:

\begin{equation}
\varphi =\frac{q}{R-{\bf R}\bfbeta},\qquad
A_{i}=\varphi \beta_{i},
\end{equation}
where $q$ means electric charge, $R=\sqrt{
\sum_{i=1}^{3}(x_{i}-x_{qi})^{2}}$, $\bfbeta=\frac{
{\bf v}}{c}$,  $R=\left| {\bf R}\right|=|{\bf r}-{\bf r}_q|$, $x_i$ are
coordinates of the observation point, and ${\bf v}$ is the velocity of the
particle with coordinates $x_{qi}$ at the instant $\tpr$ (the  earlier
time for which the time of propagation of the signal from the point ${\bf
r}_q(\tpr)$, where the charge $q$ was located, to the observation point
${\bf r}$ just coincides with the difference $t-\tpr$, $\tpr$ is
determined by the equation $\tpr-t=R/c$), and that the correct way to write down the Lorentz gauge is:

\begin{equation}
\sum_{i=1}^{3}\frac{\wpart A_{i}}{\wpart
x_{i}}+\frac{1}{c}\frac{\wpart\varphi }{\wpart t}=0\qquad{\rm or}\qquad
\sum_{i=1}^3\varphi\frac{\wpart\beta_i}{\wpart x_i}+
\sum_{i=1}^3\beta_i\frac{\wpart\varphi}{\wpart x_i}+
\frac{1}{c}\frac{\wpart\varphi}{\wpart t}=0
\end{equation}
with Brownstein's {\it whole-partial} derivatives that are noted by the symbol
$\wpart$. The strategy that we shall use is to show, in section II, that a
calculation of the Lorentz gauge for the LW-potentials using the form (1)
leads to an incorrect result.  Then in section III we show that the
calculation using the equations  (3) gives the correct result. In section
IV we comment on the possibility  of saving the use of {\it usual} partial
derivatives, and we enumerate the reasons to discard it.  In section V we
present the conclusions.

\section{The LW-potentials do not satisfy the Lorentz gauge.}

It is standard knowledge that the Lorentz gauge is satisfied in general by
the retarded potentials  (solutions of unhomogeneous D'Alembert equations)
as a consequence of charge  conservation, hence, in particular for the
LW-potentials this must be the case.

However, let us show that  traditional wisdom is wrong. In order  to do it
we proceed to calculate partial derivatives of the LW-potentials:

\begin{equation}
\frac{\partial \varphi}{\partial t}=
-\frac{q}{(R-{\bf R}\bfbeta)^2}
\left(\frac{\partial R}{\partial t}-\frac{\partial}{\partial t}
({\bf R}\bfbeta)\right),
\end{equation}
obviously, $R$ does not depend explicitly on $t$ because by definition
$R=\sqrt{\sum(x_i-x_{qi})^2}$, so $\frac{\partial R}{\partial t}=0$.
However, the term $\frac{\partial}{\partial t}({\bf
R}\bfbeta)$  also is zero because $\bfbeta$ does not depend on
$t$ explicitly: the point is that the variables like $x_{qi}(\tpr)$
and $\beta_i[v_i(\tpr)]$ depend on $t,x_i$ implicitly by the function
$\tpr(x_i,t)$.

Then
\begin{equation}
\frac{\partial\varphi}{\partial t}=0.
\end{equation}
In turn
\begin{equation}
\frac{\partial\varphi}{\partial x_i}=-\frac{q}{(R-{\bf R}{\bfbeta})^2}
\left(\frac{\partial R}{\partial x_i}-\frac{\partial}{\partial x_i}
({\bf R}{\bfbeta})\right)
\end{equation}
\begin{equation}
\frac{\partial R}{\partial x_i}=\frac{x_i-x_{qi}}{R}\qquad{\rm and}\qquad
\frac{\partial}{\partial x_i}({\bf R}{\bfbeta}) =\beta_i.
\end{equation}
Therefore as a final result we get:
\begin{equation}
\frac{\partial\varphi}{\partial x_i}=-\frac{q}{(R-{\bf R}{\bfbeta})^2}
\left(\frac{x_i-x_{qi}}{R}-\beta_i\right).
\end{equation}
With the derivatives (5) and (8) we can write for the Lorentz gauge given
by Eq. (1) the following expression:
\begin{equation}
-\frac{q}{(R-{\bf R}{\bfbeta})^2}\left(\sum_{i=1}^{3}\left[
\frac{x_i-x_{qi}}{R}-\beta_i\right]\right)
\end{equation}
which {\it is not} identically zero.

An objection can be raised against our procedure of calculation: the
LW-potentials are {\it retarded potentials}, that is, functions evaluated
at the retarded time $\tpr$ which depends on variables
$x_1,x_2,x_3,t$ and which was obtained as an implicit function from the
equation:
\begin{equation}
\sum^3_{i=1}[x_i-x_{qi}(\tpr)]^2=c^2(t-\tpr)^2,
\end{equation}
therefore, the objection continues, there are missing terms and the
calculation is wrong. However, we can offer the following answer, which
picks up the crux of the problem: for the calculation of the {\it usual}
partial derivatives we must first suppose that LW-potentials are not
evaluated at the retarded time $\tpr$, only {\it after} the
calculation the retarded is introduced, in the following form:
\begin{equation}
\left.\frac{\partial\varphi}{\partial x_i}\right|_{\tpr}\quad
{\rm and\;in\;this\;way\;for\;all\;partial\;derivatives}.
\end{equation}

With the use of this rule, which is {\it consistent} with the {\it
generally accepted} concept of partial derivatives, the LW-potentials do
not satisfy the Lorentz gauge, because as we can see from the expression
(9), even if we could find the explicit form of $\tpr$, it could not
change the form of (9) in such a way to get a zero identically. We must
also note that it is not only necessary to know the form of $\tpr$,
we need the explicit path of the particle too.

Of course, the idea to consider the potentials as functions not evaluated
at the retarded time has to seem wrong. However, it is common knowledge the
difficulty that the functions evaluated at the retarded time are not easy
to partially differentiate. Just consider the following comments in order
to sustain the assertion:
\begin{quotation}
{\sl 1. ``...in differentiating them} (the potentials) {\sl to obtain the
fields it must be noted that derivatives with respect to the position of
the field point must be taken at constant observation time, and
derivatives with respect to the observation time at fixed field points.
Since the retarded time appears explicitly in the potentials, care must be
taken to obtain the correct derivatives"} [5]
\end{quotation}

In the case of a charge in the uniform motion it is possible to eliminate
the retarded time. However, this is not an option for the case of an
non-uniformly moving point charge. So, as in the quoted cite we have to
maintain constant the field point, while the observation time does not have to be constant.

Probably W. Panofsky and M. Phillips [6] give a better statement of the
problem:

\begin{quotation}
{\sl 2. ``Partial differentiation with respect to $x_\alpha$ compares the
potentials at neighboring points at the same time, but these potential
signals originated from the charge at different times. Similarly, the
partial derivative with respect to $t$ implies constant $x_\alpha$, and
hence refers to the comparison of potentials at a given field point over
an interval of time during which the coordinates of the source will have
changed}."
\end{quotation}

The solution which was used by W. Panofsky and M. Phillips was clear: a
transformation of coordinates involving the transformation of the operator
$\frac{\partial}{\partial t}$, the generator of the time translations. So,
they changed the coordinates to a system where the concept of partial
differentiation can be applied, however, the quoted cite shows that the
real problem is the use of the concept of partial differentiation defined
on a given {\it coordinate cover} of the underlying manifold. Why? Because
as it has been stated, a function  of the form
$f\{x_i,x_{qi}[\tpr(x_i,t)]\}$ {\it cannot be} partially
differentiated with respect to spatial coordinates. Indeed, how one can
maintain constant the function $\tpr(x_i,t)$ while the $x_i$ are
varying? The answer is clear: in the fixed coordinate cover this is {\it
not possible}!

The potentials are retarded functions, so, it is not possible the use of a
{\it usual} partial derivative in a direct manner as it is indicated in
the usual statement of the Lorentz gauge (1) because a
clear contradiction with the usual concept of the  partial
differentiation is involved. So a solution must be proposed, and this solution must
take into account the fact that a {\it real} contradiction is involved
with the  concept of the {\it partial differentiation}.

To this point, we have showed that in the usual coordinate cover in which
the potentials $A_i,\varphi$ are written  the Lorentz gauge equation (1) is
not satisfied, and the reason for this must be clear: a function of the
form $f\{x_i,x_{qi}[\tpr(x_i,t)]\}$ has been subjected to a
coordinate transformation involving the retarded time $\tpr$, and the
partial derivative {\it is not an invariant concept} of the underlying
manifold, so Maxwell's equations {\it are not invariant} in front of the
general group of diffeomorfisms. Of course,  the retarded time is
an essential ingredient in the LW-potentials, so we must always take it
into account but in the proper manner: recognizing the limitation in the
use of partial derivatives and generalizing the expression (1).

Then we have come to these results:
\begin{quotation}
A. The partial differentiation of functions like the retarded potentials
is not possible, because the partial differentiation is not an invariant
operator.

B. When it is possible it is generally wrong. One can consider that the
use of just a $t$ and not a $\tpr$ in the functions $A_i,\varphi$ is
an approximation: $t\approx \tpr$. The well-known dipole
approximation [7] is $\tpr\approx t+\frac{\sqrt{\sum x_i^2(t)}}{c}$
and it is equally wrong.  \end{quotation}

As we shall show in the next section the adequate way to deal with the
transformation of involved coordinates is the use of the ``whole-partial
derivative operator". In this case a function like
$f\{x_i,x_{qi}[\tpr(x_i,t)]\}$ has the following whole-partial
derivatives:
\begin{equation}
\frac{\wpart f}{\wpart t}=\left.\frac{\partial f}{\partial
t}\right|_{\tpr}+
\sum_i\left.\frac{\partial f}{\partial x_{qi}}
\right|_{\tpr}\left.\frac{dx_{qi}}{d \tpr}\right|_{\tpr}
\left.\frac{\partial \tpr}{\partial t}\right|_{\tpr},
\end{equation}

\begin{equation}
\frac{\wpart f}{\wpart x_i}=\left.\frac{\partial f}{\partial
x_i}\right|_{\tpr}+
\sum_i\left.\frac{\partial f}{\partial x_{qi}}
\right|_{\tpr}\left.\frac{dx_{qi}}{d\tpr}\right|_{\tpr}
\left.\frac{\partial \tpr}{\partial x_i}\right|_{\tpr},
\end{equation}
where the involved partial derivatives are common partial derivatives
taken with the function $f$ independent of the retarded time $\tpr$ and,
once taken the derivative, evaluated at the retarded time. Of course, we
can see that if the retarded time $\tpr$ is not a function of the
coordinates and time $t$ we get the results : $\frac{\wpart f}{\wpart
t}=\frac{\partial f}{\partial t}$, $\frac{\wpart f}{\wpart
x_i}=\frac{\partial f}{\partial x_i}$, so we have a genuine
generalization.

\section{The LW-potentials satisfy a generalized Lorentz gauge}

Against the previous background we can try a different way to calculate
the Lorentz gauge for the LW-potentials.

First of all we must consider them as retarded functions, that is
functions evaluated at the retarded time $\tpr$ {\it before} the process
of differentiation takes place. So, we use {\it whole-partial derivatives}
to made the calculations. In this way we shall show that  LW-potentials
satisfy the Lorentz gauge as given by Eq. (3), as must be the case if the
rule of  generalization used with the Maxwell's equations is right. Now,
let's start to calculate the expression (3) (the symbols $\wpart$
denote the Brownstein's whole-partial derivative, the symbols $d$ denote
the usual whole derivative with respect to a given variable, the symbols
$\partial$ denote the usual partial derivative, $\nabla_q$ denotes the
usual operator ``nabla" with respect to $x_{qi}$, $\nabla$ denotes the
usual operator ``nabla" with respect to $x_i$ and $\nabla_v$ denotes the
usual operator ``nabla" with respect to $v_i$).  Besides we indicate the
kind of a functional dependence for each involved function:

\begin{equation}
{\bf R}={\bf R}\{x_i,x_{qi}[\tpr(x_i,t)]\};\qquad \bfbeta=
\bfbeta\{{\bf v}[\tpr(x_i,t)]\};\qquad \bfeta =\frac{d \bfbeta}{d\tpr};
\end{equation}

\begin{equation}
\frac{1}{c}\frac{\wpart\varphi}{\wpart t}=
-\frac{q}{(R-{\bf R}\bfbeta)^2}\left(\frac{1}{c}\frac{\wpart R}{\wpart t}-
\frac{1}{c}\frac{\wpart}{\wpart t}({\bf R}\bfbeta)\right),
\end{equation}
where

\begin{equation}
\frac{1}{c}\frac{\wpart R}{\wpart t}=\frac{\partial\tpr}{\partial t}
(\nabla_q R\cdot\bfbeta)=-\frac{\partial\tpr}{\partial t}
\left(\frac{{\bf R}\bfbeta}{R}\right),
\end{equation}

\begin{equation}
\frac{1}{c}\frac{\wpart}{\wpart t}({\bf R}\bfbeta)  =
\frac{\partial\tpr}{\partial t}\biggl[\nabla_q({\bf R}\bfbeta)\bfbeta +
\nabla_v({\bf R}\bfbeta)\bfeta\biggr].
\end{equation}
Now, using well-known vector operations we obtain:

\begin{equation}
\nabla_q({\bf R}\bfbeta)={\bf R}\times(\nabla_q\times\bfbeta)+
\bfbeta\times(\nabla_q\times{\bf R})+({\bf R}\cdot\nabla_q)\bfbeta+
(\bfbeta\cdot\nabla_q){\bf R}=-\bfbeta,
\end{equation}
and
\begin{equation}
\nabla_v({\bf R}\bfbeta)={\bf R}\times(\nabla_v\times\bfbeta)+
\bfbeta\times(\nabla_v\times{\bf R})+({\bf R}\cdot\nabla_v)\bfbeta+
(\bfbeta\cdot\nabla_v){\bf R}=\frac{{\bf R}}{c},
\end{equation}
so, with the help of Eqs. (16)-(19) and taking into account that (see [10])
$$
\frac{\partial\tpr}{\partial t}=\frac{R}{R-{\bf R}\bfbeta},
$$
we can write (15) as:  \begin{equation}
\frac{1}{c}\frac{\wpart\varphi}{\wpart t}=-\frac{qR}{(R-{\bf R}\bfbeta)^3}
\left(\frac{{\bf R}\bfbeta}{R}-\beta^2+\frac{{\bf R}\bfeta}{c}\right).
\end{equation}

Now:
\begin{equation}
\frac{\wpart\varphi}{\wpart x_i}=-\frac{q}{(R-{\bf R}\bfbeta)^2}
\left(\frac{\wpart R}{\wpart x_i}-\frac{\wpart}{\wpart x_i}({\bf
R}\bfbeta)\right),
\end{equation}
where

\begin{equation}
\frac{\wpart R}{\wpart x_i}=\frac{\partial R}{\partial x_i}+
(\nabla_qR\cdot{\bf v})\frac{\partial\tpr}{\partial x_i},
\end{equation}
and
\begin{equation}
\frac{\wpart}{\wpart x_i}({\bf R}\bfbeta)=
\frac{\partial}{\partial x_i}({\bf R}\bfbeta)+
\left[\nabla_q({\bf R}\bfbeta)\cdot{\bf v}+\nabla_v({\bf R}\bfbeta)\cdot
\frac{d{\bf v}}{d\tpr}\right]\frac{\partial\tpr}{\partial x_i}.
\end{equation}
This allows us, taking into account that
$$
\nabla\tpr=-\frac{{\bf
R}}{c(R-{\bf R}\bfbeta)}\quad ({\rm see}\; {\rm Ref.}\; 10,\; {\rm Eq.}\;
63.7),
$$
to write down  (after a straightforward calculation):
\begin{equation} \sum_{i=1}^3\beta_i\frac{\wpart\varphi}{\wpart x_i}=
-\frac{q}{(R-{\bf R}\bfbeta)^3}\left[\left(1+\frac{{\bf
R}\bfeta}{c}\right){\bf R}\bfbeta-\beta^2R\right].
\end{equation}
The last term to be calculated in Eq. (3) is
\begin{equation}
\sum_{i=1}^3\frac{\wpart\beta_i}{\wpart x_i}=\nabla\tpr\cdot\bfeta,
\end{equation}
so
\begin{equation}
\sum_{i=1}^3\varphi\frac{\wpart\beta_i}{\wpart x_i}=
-\frac{q}{(R-{\bf R}\bfbeta)^2}\left(\frac{{\bf R}\bfeta}{c}\right).
\end{equation}
Now let us substitute the results (20), (24) and (26) into {\it lhs} of
(3):
\begin{equation}
\sum_{i=1}^3\varphi\frac{\wpart\beta_i}{\wpart x_i}+
\sum_{i=1}^3\beta_i\frac{\wpart\varphi}{\wpart x_i}+
\frac{1}{c}\frac{\wpart\varphi}{\wpart t},
\end{equation}
and after a straightforward calculation one can easily make sure that
the expression (27) is identically {\it zero}.

It means that the Lorentz gauge for the  LW-potentials is satisfied if we
use Brownstein's {\it whole-partial} derivatives.

\section{Diverse interpretations}

There is a way of saving the use of partial differentiation, but nevertheless
the Lorentz gauge must change.

If we have the retarded potentials given as
\begin{equation}
\varphi=\varphi(x_k,t,x_{qk},\tpr,v_k), k=1,2,3,
\end{equation}
\begin{equation}
A_i=A_i(x_k,t,x_{qk},\tpr,v_k),i=1,2,3,
\end{equation}
then if we know the functional dependence of each argument we can use an
adequate process of {\it whole-partial} differentiation, as it was done in
section III, but if we know the function $\tpr$ as an explicit function of
coordinates and time $t$, then we can write the following functions (after
substitution of the explicit form of the function $\tpr=f(x,y,z,t)$ in Eqs.
(28), (29):
\begin{equation}
\varphi=\psi(x,y,z,t),
\end{equation}
\begin{equation}
A_i=a_i(x,y,z,t),
\end{equation}
where $\psi$ and $a_i$ are explicit function of the coordinates and time.
Obviously, if we have the explicit functions $\psi$ and $a_i$, we must
have the following equalities:
\begin{equation}
\frac{\partial\psi}{\partial t}=\frac{\wpart\varphi}{\wpart t},
\end{equation}
\begin{equation}
\frac{\partial a_i}{\partial x_i}=\frac{\wpart A_i}{\wpart x_i},
\end{equation}
but we cannot use, in an interpretation of this kind which tries to save
the use of partial derivatives, the {\it rhs} of Eqs. (32), (33),
because then the interpretation would  not be an independent
interpretation. The point is that  in this interpretation we must try to
find the explicit form of $\tpr$, and to give an explicit path. This is
the only way in which we have really an independently defined
interpretation which saves the use of partial derivatives.

Now, in order to get the explicit form of $\tpr$ we have to integrate the
following differential form:
\begin{equation}
\frac{\partial\tpr}{\partial t}dt+\nabla\tpr\cdot d{\bf r}
\end{equation}
which does not seem like an easy task, because in fact, the differential
form (34) is not an integrable 1-form. We can check this assertion
calculating the {\it cross partial derivatives} as follows\footnote{Why we must use partial derivatives? If the coefficients
involve the retarded time this cannot be the case. However, the
coefficients of the differential 1-form (34) does not involve the
evaluation at the retarded time $\tpr$ because the functional dependence
of the retarded time  is given by $\tpr= f(x,y,z,t)$ a functional
dependence not involving the evaluation of $f$ (a function obtained by the
implicit function theorem applied to $\tpr=t-\frac{R(\tpr)}{c}$) at any
retarded time.}:

\begin{equation}
\frac{\partial}{\partial x_i}\left(\frac{R}{R-{\bf R}\bfbeta}\right)=
\left(-\frac{x_i-x_{qi}}{R}\right)\left(\frac{1}{R-{\bf R}\bfbeta}\right)-
\left(\frac{R}{(R-{\bf R}\bfbeta)^2}\right)\left(-\frac{x_i-x_{qi}}{R}-
\beta_i\right),
\end{equation}
\begin{equation}
\frac{\partial}{\partial t}\left(\frac{x_i-x_{qi}}{c(R-{\bf
R}\bfbeta)}\right)=\left(\frac{\beta_i}{R-{\bf R}\bfbeta}\right)+
\left(\frac{x_i-x_{qi}}{c(R-{\bf R}\bfbeta)^2}\right){\bf R}\bfeta.
\end{equation}
So one can see  the non-integrability of (34). In this case the general
integral of the equation  cannot be expressed using some single function,
instead, in general, we require several ones.

Another way to get $\tpr$ relies on two points:

- we explicitly  know the path of the particles;

- if the prior point is given, then we can write $\tpr$ as
$\tpr=t-\frac{R(\tpr)}{c}$ and we can use the well-known Lagrange
development to get

\begin{equation}
\tpr=t-\sum_{n=1}^\infty\frac{c^{-n}}{n!}\frac{d^{n-1}R(t)}{dt^{n-1}}
\end{equation}
and test its convergence behavior. Clearly $c^{-n}$ is decreasing, but we
need to know if the trajectory is bounded or not. Yet we can find another
way to save the use of partial derivatives in the correspondent
literature, but introducing a new set of concepts (see Ref [8] pp. 47 and
69), for example, Faraday's law is correctly written with partial
derivatives if we leave aside the vector character of the magnetic field
${\bf B}$ and introduce a two-form $B_{\lambda\mu}dx_\lambda\wedge
dx_\mu$, and then we write Faraday's law as (Post's notation):
\begin{equation}
2\partial_{[\nu}E_{\lambda]}=-\frac{\partial}{\partial t}B_{\nu\lambda}.
\end{equation}
As it was emphasized by Post, the transformation behavior of $E_\lambda$ and
$B_{\nu\lambda}$ comes from the transformation properties of the tensor
$F_{\mu\nu}$ (which is the usual electromagnetic tensor) under holonomic
transformations (see Ref. [8] pp. 57-58). Clearly Eq. (38) is not the
usual Faraday's law, but it is a generalized form.

However, if we retain the vector character we need the use of {\it
whole-partial} derivatives and we can guess why: by definition partial
derivatives cannot take into account transformation properties, and when
we evaluate functions at the retarded time $\tpr$ we are making an
non-holonomic transformation\footnote{Non-holonomic because we cannot integrate the expression (34)
to get an explicit function $\tpr=f(x,y,z,t)$. And in this case Post's
proof of the invariance of the Maxwell's equations given by
$\partial_{[\kappa}F_{\lambda\mu]}=0$, $\partial_\nu G^{\lambda\nu}= 4\pi
c^\lambda$ fails because it is only valid for holonomic transformations
(see Ref. [8] pp. 57-61).}. We can explain this fact in more
detail:

Let us consider the operator
\begin{equation}
X=\sum_{i=1}^n\xi_i\frac{\partial}{\partial x_i}
\end{equation}
under the holonomic transformation:
$\overline{x_i}=\overline{x_i}(x_1,\ldots,x_n), i=1,\ldots,n$. It is
a standard knowledge that the transformation rule for (39) is given by

\begin{equation}
\sum_i\xi_i(x)\frac{\partial}{\partial x_i}=
\sum_{i,k}\xi_i[x(\overline{x})]\frac{\partial\overline{x_k}}{\partial
x_i}\frac{\partial}{\partial\overline{x_k}}=\overline{X}.
\end{equation}
Now it is clear that (39) is just a {\it whole-partial} derivative\footnote{A {\it whole-partial} derivative is just a case of a {\it
total} derivative. Just consider the operator (39), if we choose the
coefficients $\xi_i$ in an adequate manner we can get, without doubt, the
expressions (12), (13).  It is possible to choose the invariant parameter
by just choosing a coefficient: for some $i$ we put $\xi_i=1$. It is clear
that the concept of a {\it whole-partial derivative} has some ground
because the functions involve several variables, but the basic concept is
that of a ``total derivative", as two authors of the present work have
tried to show in Ref. [4].}
$\frac{\wpart}{\wpart s}$ if the integral curves of the vector field (39)
are given by $\frac{\wpart x_i}{\wpart s}=\xi_i(x)$. Under the holonomic
transformation used, which does not affect the arbitrary parameter $s$,
the integral curves of the transformed vector field are
$\frac{\wpart\overline{x_i}}{\wpart
s}=\sum_k\xi_i[x(\overline{x})]\frac{\partial\overline{x_k}}{\partial
x_i}$ but we still have just a {\it whole-partial} derivative
$\frac{\wpart}{\wpart s}$ which remains invariant under the transformation.
This is the reason, the underlying mathematical reason of the
generalization realized in Ref. [1]. Besides, it is clear that the use of
the transformation rule (40) is not limited to holonomic
transformation\footnote{It is possible to say that the differential 1-form (34) is
{\it of necessity}  the form $d\tpr$. However, the criterion for
decision if (34) is an integrable 1-form or not is only given by cross
differentiation.
}, instead we  can use transformations of the form
$\sum_kA_i^k\xi_i$ where $A_i^k$ does not arise from a holonomic
transformation but, however, are non-singular transformations. So, in
this sense, when we change our coordinates to coordinates including the
retarded time $\tpr$ we are changing to what is known not only as a
non-holonomic reference frame, but as a set of quasi-parameters of the
kind often found in dynamics [11].

\section{Conclusions}

We have showed that the correct way to write down the Lorentz
gauge is

\begin{equation}
\sum_{i=1}^3\frac{\wpart A_i}{\wpart
x_i}+\frac{1}{c}\frac{\wpart\varphi}{\wpart t}=0
\end{equation}
for the case of the LW-potentials. Hence previous generalization (see Ref.
[1]) of the Maxwell's equations is really a generalized form and we must
take Eq. (3) as the general way to write down the Lorentz gauge and, of
course, all gauges.

The underlying reasoning is clear: if the form of writing (1) would be
general, it should be valid for {\it all the cases}. However we have shown
at least one exception, then Eq. (1) {\it is not general}. And as we have
shown the Brownstein's generalization (3) gives the correct results.

We have discussed the possibilities to continue using  partial
derivatives, we have deduced that for that interpretation to be useful one
requires either: (i) an explicitly defined functional form for the retarded
time $\tpr$ or (ii) according to Post change the vector formalism for the
tensor formalism (change ${\bf B}$ for $B_{\lambda\nu}$) which is another
kind of generalization.

However we can use the Post's propositions\footnote{Let's reproduce them here: 1) ``Physicists cannot think
without a cartesian frame", 2) ``Physicists cannot think without explicit
reference to an inertial frame" (see Ref. [9] p. 77).} to reject the use of
partial derivatives: partial derivatives are always well-defined in a
given set of coordinates, however, when we use changes  of coordinates we
use implicitly the transformation rule (40), which is just of a {\it
whole-partial} derivative.

\acknowledgments

The authors would like to express their gratitude to Prof. Valeri
Dvoeglazov for his discussions and critical comments. We would also like
to thank Ernesto Mendivil for revising the manuscript.

\end{document}